\documentclass[12pt,a4paper]{article}
\usepackage{amsmath}
\usepackage{graphicx}
\usepackage{times}
\usepackage{cite}
\begin{document}
\begin{titlepage}
\title{Increasing effective intensity of soft strong interactions}
\author{ S.M. Troshin, N.E. Tyurin\\[1ex]
\small   NRC ``Kurchatov Institute''-- IHEP\\
\small   Protvino, 142281, Russian Federation}
\normalsize
\date{}
\maketitle

\begin{abstract}
We suggest  definition of  effective interaction intensity for soft hadron collisions and  discuss   its  energy dependence in the preasymptotic region.  Practical importance of this quantity consists in separation   of rising interaction radius from effective interaction intensity increase both contributing to the total cross--section growth. It would be helpful for understanding the origin of this growth at the accelerator energies. The essential feature is that the effective interaction intensity is an experimentally measurable quantity.
\end{abstract}
\end{titlepage}
\setcounter{page}{2}
\section*{Introduction. The interaction intensity}
In general, the interaction intensity is determined by the magnitude of coupling constant in the respective Lagrangian.
It is  a notorious  statement that the current theory of strong interactions, QCD, does not allow  one to obtain a solution  in the nonperturbative region. As a consequence an explicit relation of the coupling constant of the QCD Lagrangian with a  strength of interaction  is not known in the soft region despite a significant progress in numerical calculations in the framework of  lattice QCD.  

The hadron's dynamics at short distances is a different story. The parton model  and perturbative QCD work well in this region due to  asymptotic freedom and cross--sections of hard hadron interactions are explicitly  determined by the QCD running coupling constant $\alpha_s$.

Thus, the knowledge of the interaction dynamics of the soft hadron collisions remains  a rather limited for a long time. The range of tools is rather narrow:  general constraints of  quantum field theory  resulting from such principles as analyticity and unitarity. Those can be amended with various model assumptions and calculations based on them and/or inspired by QCD. 
An effective  interaction intensity is needed for a description of soft collisions and it is suggested to consider for that role the function $Y(s)$ defined as   a well--known ratio 
\begin{equation}\label{y}
Y(s)\equiv\sigma_{tot}(s)/16\pi B(s),
\end{equation}
where $\sigma_{tot}(s)$ is an {\it effective} total cross-section  and $B(s)$ is a slope parameter of the forward elastic proton--proton scttering. Its standard definition is:
\begin{equation}
B(s) \equiv \frac{d}{dt}\ln \frac{d\sigma}{dt}|_{t=0}
\end{equation}
 with $d\sigma/dt $ being a differential cross--section of proton--proton elastic scattering.

Eq. (\ref{y}) will be treated as a dimensionless measure of effective intensity of soft hadron interactions which is determined by  the total strength of  interaction (the total cross--section) divided by the characteristic of area defined by the radius of interaction.

The effective interaction intensity definition Eq. (\ref{y}) is based on  interpretation of the scattering in the impact parameter representation and can be considered as an effective analog of the QCD coupling constant in the soft region. The observed experimental trends and unitarity of the scattering matrix give an increasing magnitude of the effective interaction intensity in the preasymptotic energy region with $Y(s) = O(1)$ at $s\to\infty$.  It does not allow one to use  perturbation calculations for the scattering amplitude $F(s,t)$ in the form of   expansion over a parameter proportional to $Y(s)$.  Such an expansion would make sense in the case of a constant total cross--section and increasing interaction radius. It is not the case.

The impact parameter based interpretation is discussed in section 2, section 1 describes some qualitative hints on the asymptotics of the soft strong interactions which could be motivated by  the nonperturbative QCD, section 3 is devoted to the relation of the effective interaction intensity  with regime of the unitarity saturation at asymptotics (cf. e.g. \cite{bar,wall,ben}) and section 4 discusses separation of interaction radius and intensity roles in the total--cross section growth.

We address also the two other  functions $X(s)$ and $Z(s)$:
\begin{equation}\label{x}
X(s)\equiv\sigma_{el}(s)/\sigma_{tot}(s),
\end{equation}
\begin{equation}\label{z}
Z(s)\equiv X(s)Y(s)=\sigma_{el}(s)/16\pi B(s).
\end{equation}

 The first one is a relative fraction of elastic events and the function $Z(s)$  is determined as a product of the  $X$ and $Y$. Thus, all three functions are the  ratios of the experimentally measurable quantities and are a priori useful. It is very common to use $X(s)$ to distinguish between different asymptotics \cite{new}. We would like to argue that the other two function $Y(s)$ and $Z(s)$ may be even more appropriate to judge on the asymptotics as well as on  dynamics of the soft hadron interactions.

\section{QCD--inspired asymptotics and $Y(s)$.}
Our consideration is a qualitative one and, since the high energy experimental data are in favor of a pure 
imaginary amplitude,  we suppose that the real part of the elastic scattering amplitude   can be safely neglected. Thus, the $f(s,b)$  stands for an imaginary part of a scattering amplitude in the impact parameter representation.

Hints for   possible asymptotic values of the function $Y(s)$ at $s\to\infty$ can be invoked from  general features of QCD. 
   There are two important phenomena in this regime of QCD: confinement and spontaneous chiral symmetry breaking with the respective scales $\Lambda_{QCD}$ and  $\Lambda_{\chi}$. 

First, there is an option   that the above two scales coincide. It  has led to  prediction that deconfinement  would  result in production of quark-gluon plasma with  free current quarks and gluons as the degrees of freedom\cite{colper}. It is natural to expect in this case that the high-energy asymptotics of this hadron interactions would be incoherent, corresponding to the dominance of inelastic dynamics and  the black--disc limit  saturation at $s\to\infty$  and $Y(s) \to 1/2$ in this case. The origin of elastic scattering in this case would be just a shadow of all the inelastic  interactions (the absorptive scattering mode). This regime  arises due to primarily  introduced restriction for the partial amplitude $|f|\leq 1/2$ instead of the unitarity limit $|f|\leq 1$.

   However, these two scales $\Lambda_{QCD}$ and  $\Lambda_{\chi}$ may have different values \cite{chern} and spontaneous chiral symmetry breaking becomes responsible for  formation of a core in a hadron. A  proton structure can be imagined then as a hard  ball placed in the hadron central region coated by a thick but fragile peripheral matter \cite{epl}. The hypothesis on existence of a proton core  \cite{orear,frank,islam,drem} is relevant for the asymptotic  geometric elastic interactions while the interactions of a fragile peripheral matter are responsible  for  inelastic collisions and are relatively dumped. From these two, the former interaction is expected to correspond to the coherent dynamics while the latter one--- to the incoherent dynamics. Thus, the resulting dynamics will be a complicated mixture of coherent and incoherent interactions.

    A substitution for a fully incoherent inelastic dynamics, is given by the hypothesis of maximal importance of a geometric elastic scattering.  It is based on saturation of unitarity due to a maximal strength of strong interactions (i.e. maximality of the imaginary part of the elastic partial amplitude consistent with unitarity). This principle has been developed by Chew and Frautchi   along with the ``strip approximation'' \cite{ch} and is associated with a coherent dynamics of hadron interactions. The ratio of the rates of elastic only to elastic plus inelastic collisions has asympotic value of unity, i.e.  $\sigma_{el}/\sigma_{tot}\to 1$  at $s \to\infty$. This is  called the reflective scattering mode ($f>1/2$) and correlation of elastic scattering with inelastic processes is negative (antishadowing) in this mode.

    A few  additional notes are needed for clarification of the main distinctions of the reflective and  absorptive scattering modes. 
    The unitarity  allows variation of the function $f$  in the interval
    $0\leq f \leq 1$. The absorptive scattering mode assumes that the region of elastic scattering amplitude  variation is reduced to $0\leq f \leq 1/2$. The value of  $f=1/2$ corresponds to a complete absorption of an initial state, i.e. respective elastic scattering element of the scattering matrix becomes equal to zero, $S=0$  ($S=1-2f$). It is  recognised  that this limit (black disc limit), in fact, has been reached at
    small $b\sim 0$ values of the impact parameters at the LHC energies. The distinctive feature of a transition to the reflective scattering mode is exceeding  value $1/2$ with simultaneous development   of the  peripheral form (with minimum at $b=0$) for the inelastic overlap function (see \cite{new} and references therein).

    We emphasize here the role of the function $Y(s)$ in the detection of a possible presence of the reflective scattering mode and discuss  energy dependence of  $Y(s)$ with regard of its role as an intensity  of soft hadron interactions. 
   The advantage of this definition of an effective intensity has been  mentioned:  $Y(s)$ can be experimentally measured and  its energy dependence of  has been studied  for a long  time.  An increasing energy dependence of  $Y(s)$ has been observed at high energies (cf. \cite{dmdpi}).
   
   To illustrate the notion of the effective interaction intensity, it is instrumental consider an oversimplified   scattering amplitude in the form of a step function. When  the scattering amplitude covers a rectangle area in the
    impact parameter  representation the function $Y(s)$ is just a height of the rectangle (Fig. 1).
   \begin{figure}[hbt]
   	\begin{center}
   		\resizebox{10cm}{!}{\includegraphics*{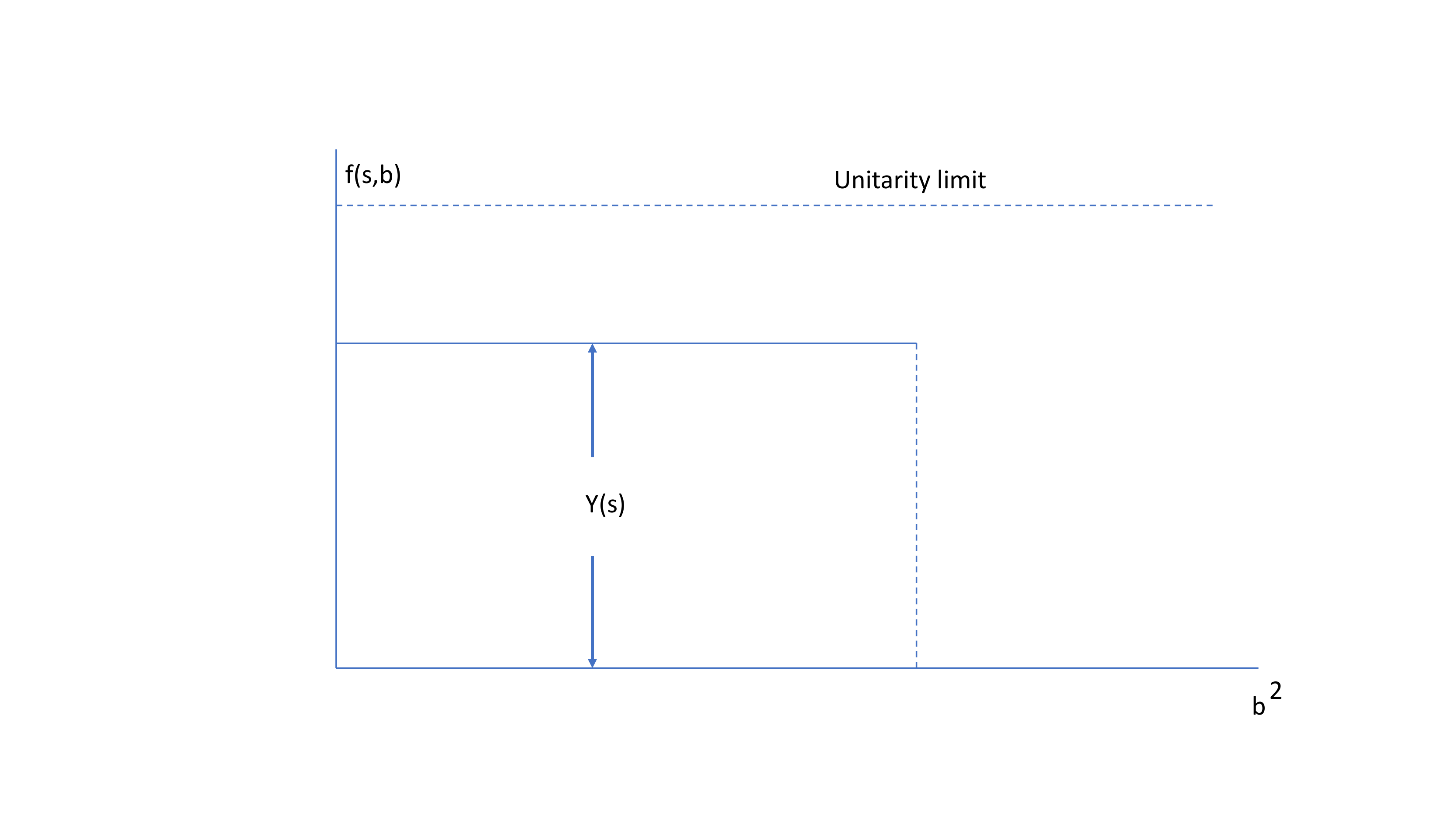}}
   	\end{center}
   	\vspace{-1cm}
   	\caption[ch3]{\small Illustration of the effective interaction intensity notion.}
   \end{figure}
   At available energies, such form of the scattering amplitude is more relevant for interactions of nuclei, but not  hadrons. Collisions of large objects like nuclei demonstrate clear pattern of diffraction with many dips and bumps in the elastic differential cross--section  since matter distribution in nuclei can be well approximated by Fermi function. 
   
   Contrary to a nuclear case, hadronic amplitudes  in the impact parameter representation can be better approximated by a Gaussian function.
   Such approximation corresponds to an exponential decrease with $-t$ of the differential cross--section. 
  
  Future experiments  at high energies would be crucial for the studies of hadron dynamics in the nonperturbative sector of QCD and allow  one to perform a reasonable  selection  among the principles of  primary role of particle production and  maximal strength of strong interactions  in the limit $s\to\infty$. 

Currently, an effective interaction intensity $Y(s)$ depends on energy in a nontrivial way and is far from being constant.   A nontrivial energy dependence  is just another indication of the remotness of the asymptotics. 

Thus, the above said is intended to emphasize the role of effective interaction intensity for the determination of the origin of dynamics at asymptotics.

\section{Effective interaction intensity}
Unitarity relation  can be written in the following form:
\begin{equation}
h_{tot}(s,b)=h_{el}(s,b)+ h_{inel}(s,b),
\end{equation}
where the overlap functions $h_{i}(s,b)$ ($i=tot,el,inel$) are the total contributions of all, elastic or inelastic intermediate states into the unitarity relation, respectively. The overlap functions $H_i(s,t)$ are the properly normalized Fourier--Bessel transforms of the functions $h_i(s,b)$ and can be written at small values of $|t|$ in the form of expansion over transferred momentum \cite{adj}:
\begin{equation}
H_i(s,t)=\sigma_i(s)[1+\frac{1}{4}\langle b^2\rangle_i(s) t+...],
\end{equation}

Study of geometrical properties of hadron interactions    represents an important step \cite{blg} toward  understanding of hadron dynamics related to the  development of QCD in the nonperturbative region.  Those properties are encoded in the impact--parameter dependencies of the total, elastic and inelastic overlap functions. Knowledge of these functions provides more information on hadron interactions compared to the integrated over the impact parameter observables. Despite the impact parameter analysis indicated an existence of the reflective scattering mode at the LHC (when the scattering amplitude $f$ in the impact parameter representation is greater than 1/2) this conclusion is not currently commonly accepted. The reason is that an impact parameter analysis invokes several additional assumptions. The question: which limit for the scattering amplitude (if any) is saturated at $s\to\infty$, the black disc  or the unitarity limit is considered still unanswered.

The expectation values $\langle b^2\rangle_i(s)$ (forward slopes of the overlap functions) are the measures of the reaction perepherality or centrality
\begin{equation}
\langle b^2\rangle_{i}(s)=\frac{\int_0^\infty b^2 h_i(s,b)bdb}{\int_0^\infty h_i(s,b)bdb},
\end{equation}
 they are interrelated  due to unitarity \cite{adj}
\begin{equation}
\langle b^2\rangle_{tot}(s)=\frac{\sigma_{el}(s)}{\sigma_{tot}(s)}\langle b^2\rangle_{el}(s)+
\frac{\sigma_{inel}(s)}{\sigma_{tot}(s)}\langle b^2\rangle_{inel}(s).
\end{equation}
Numerical estimates for those quantities have been given in \cite{des}. 
It was shown  that the range of hadronic forces responsible for elastic and inelastic interactions are 
around 0.8 and 1.3 fm at the LHC energies, respectively. Their energy dependencies relevant for the absorptive and reflective scattering modes have been considered in \cite{mpla}. There are  relations among root--mean--squares of these quantities valid for a wide range of energies and verified at the CERN ISR, S$\bar {p}p$S and LHC:
\begin{equation}
\sqrt{\langle b^2\rangle_{el}(s)}<\sqrt{\langle b^2\rangle_{tot}(s)}<\sqrt{\langle b^2\rangle_{inel}(s)}.
\end{equation}

Note, that for the reflective scattering mode the two interaction radii concide asymptotically
\begin{equation}\label{bas}
\langle b^2\rangle_{tot}(s) \to \langle b^2\rangle_{el}(s)
\end{equation}
at $s\to\infty$ since $\sigma_{el}(s)/\sigma_{tot}(s)\to 1$. It means that the phase of the scattering amplitude in this limit corresponds to the case of a pure imaginary amplitude.

Since the slope parameter $B_{tot}(s)=\langle b^2\rangle_{tot}(s)/2$, the function $Y(s)$ can be presented in the following form:
\begin{equation}\label{defy}
Y(s)=\int_0^\infty f(s,b)bdb/\langle b^2\rangle_{tot}(s).
\end{equation}
Eq. (\ref{defy}) further clarifies  the term interaction intensity. When the amplitude $f(s,b)$ is a decreasing function of $b$ with maximum at $b=0$ the following inequality is valid
\begin{equation}\label{ineq}
Y(s)<f(s,b=0).
\end{equation}
The clues on the energy dependence of the function $f(s,b=0)$ can be obtained under studies of the $d\sigma/dt$ in the deep--elastic region \cite{mpla}.

  It should be noted that transition to the reflective scattering mode ($S<0$) makes  s and b--dependencies of the elastic overlap function  steeper than the dependencies of the scattering amplitude $f(s,b)$ because of the relations for   derivatives \cite{phrev}:
 \begin{equation}
 \frac{\partial h_{el}(s,b)}{\partial s}=[1-S(s,b)]\frac{\partial f(s,b)}{\partial s},
  \end{equation}
 \begin{equation}
\frac{\partial h_{el}(s,b)}{\partial b}=[1-S(s,b)]\frac{\partial f(s,b)}{\partial b}.
\end{equation}
 The above variations of the energy and impact parameter dependencies are due to the reflective mode appearance, i.e. transition to the values of the factor 
$1-S$ from the range where $1-S<1$ to the range where $1-S>1$.  Such transition could explain increase of the ratio $\sigma_{el}/\sigma_{tot}$ at the LHC energies as a consequence of  redistribution of the total probability between  elastic and inelastic interactions according to unitarity \cite{rin}.

Of course, numerical reconstruction of the  overlap functions $h_i(s,b)$ is the most   sensitive way for a detection  of the reflective scattering mode presence at the LHC. However, an effective use of this reconstruction is complicated by  necessity of invoking several extra assumptions. 
To provide a more convincing answer  on the presence of the reflective scattering mode one needs to present further arguments based on the behavior of the directly measurable observables \cite{new}. These are  absence of the secondary dips in the differential cross-section of the elastic $pp$--scattering, slowdown of the mean multiplicity and increase of the average transverse momentum of  secondary particles. 

The knowledge of the functions $Y(s)$ and $dY(s)/ds$ can also be helpful for the above purpose. The dimensionless function $Y(s)$ being considered as effective interaction intensity has a different physical meaning than  $X(s)$ which is a ratio of elastic events' rate to the total collision events' rate. 
The available data and energy extrapolations for  $X(s)$  have been considered  in \cite{fag,fm}. Unfortunately, present accelerator energies are not high enougth to provide a definite conclusions on its asymptotic value.
At present energies, the both functions $X(s)$ and $Y(s)$ are increasing  with energy and have the value about $0.3$ which is even lower than their black disc limiting value of $1/2$.

Upper limits for the functions $X(s)$ and $Y(s)$ provided by unitarity, i.e.
\begin{equation} \label{mcdeeu}
X(s)\leq 1, Y(s)\leq 1.
\end{equation}
Those inequalities, Eq. (\ref{mcdeeu}), are saturated in the reflective scattering mode. Physical interpretation of this saturation is related to the relative dominance of the elastic scattering which can be based on a presence of a hard  core in the hadron structure \cite{epl}. 

The third experimentally measurable function $Z(s)$ is also limited from above by unity.
Due to relation for the derivatives
\begin{equation}
dZ(s)/ds=Y(s)dX(s)/ds+X(s)dY(s)/ds
\end{equation}
the following simple approximate relation of $dZ(s)/ds$ with
$Y(s)$ and $dY(s)/ds$ takes place\footnote{This relation presupposes that $X(s)\simeq Y(s)$.}:
\begin{equation}
dZ(s)/ds\simeq 2Y(s)dY(s)/ds.
\end{equation}
A numerical value of $Z$ is rather small at the LHC energy $\sqrt{s}=13$ TeV, it is around 0.09, and respective relation between derivatives of the functions $Z$ and $Y$ (with use of  the TOTEM  data  \cite{totem}) is
\begin{equation}
dZ(s)/ds\simeq 0.6dY(s)/ds.
\end{equation}
 The function $Z(s)$ would have a significantly steeper energy increase than the functions $X(s)$ and $Y(s)$ in case  the relective scattering mode  is realized at $s\to\infty$.
In the impact parameter representation
\begin{equation}\label{zs}
Z(s)=\int_0^\infty f^2(s,b)bdb/\langle b^2\rangle_{tot}(s).
\end{equation}
Eq. (\ref{zs}) can be interpreted as an integrated strength of the elastic interactions averaged over size of area of all the interactions. However, in the reflective scattering mode in the limit of $s\to\infty$, the function 
$Z(s)$ becomes the  intensity of elastic interactions due to Eq. (\ref{bas}). It tends to unity at $s\to\infty$ in this mode, i.e.   $$Y(s)-Z(s)\to 0$$ at $s\to\infty$ as a result of the inelastic channels  self-damping. 

Thus, this section further clarifies meaning of the effective interaction intensity based on the geometrical intuition on the interaction in the impact parameter space.

\section{ Consequences of reflective scattering}
The values of the inelastic overlap function $h_{inel}$    are very close to its limiting value ($h^{max}_{inel}=1/4$ ) in a rather broad region of the impact parameter values, i.e. from $0$ till $b\simeq 0.4$ fm at $\sqrt{s}=13$ TeV \cite{alkin1,tsrg}. The approximate unitarity  at this particular energy and  impact parameter values can be written (neglecting small real part of the scattering amplitude as usual) as
\begin{equation}
(f-1/2)^2=1/4-h_{inel}\simeq 0
\end{equation}
The extrapolations of the scattering amplitude at small impact parameters to higher values of energy   can be divergent. The one option (which assumes that elastic scattering at high energy is a pure shadow of absorption) supposes  that the value of $S=0$  would be frozen at $b=0$ and further energy increase would make amplitude  $f(s,b)$   flat and wider while keeping the same maximal value of $1/2$. This option disregards self--damping of the inelastic channels. Another possibility is that the scattering amplitude  would cross  the value of $1/2$  under the energy increase  and saturate the unitarity limit of $1$ at $s\to\infty$ becoming flat. The two possibilities  are presented schematically at Fig. 2.
 \begin{figure}[hbt]
	\begin{center}
		\resizebox{12cm}{!}{\includegraphics*{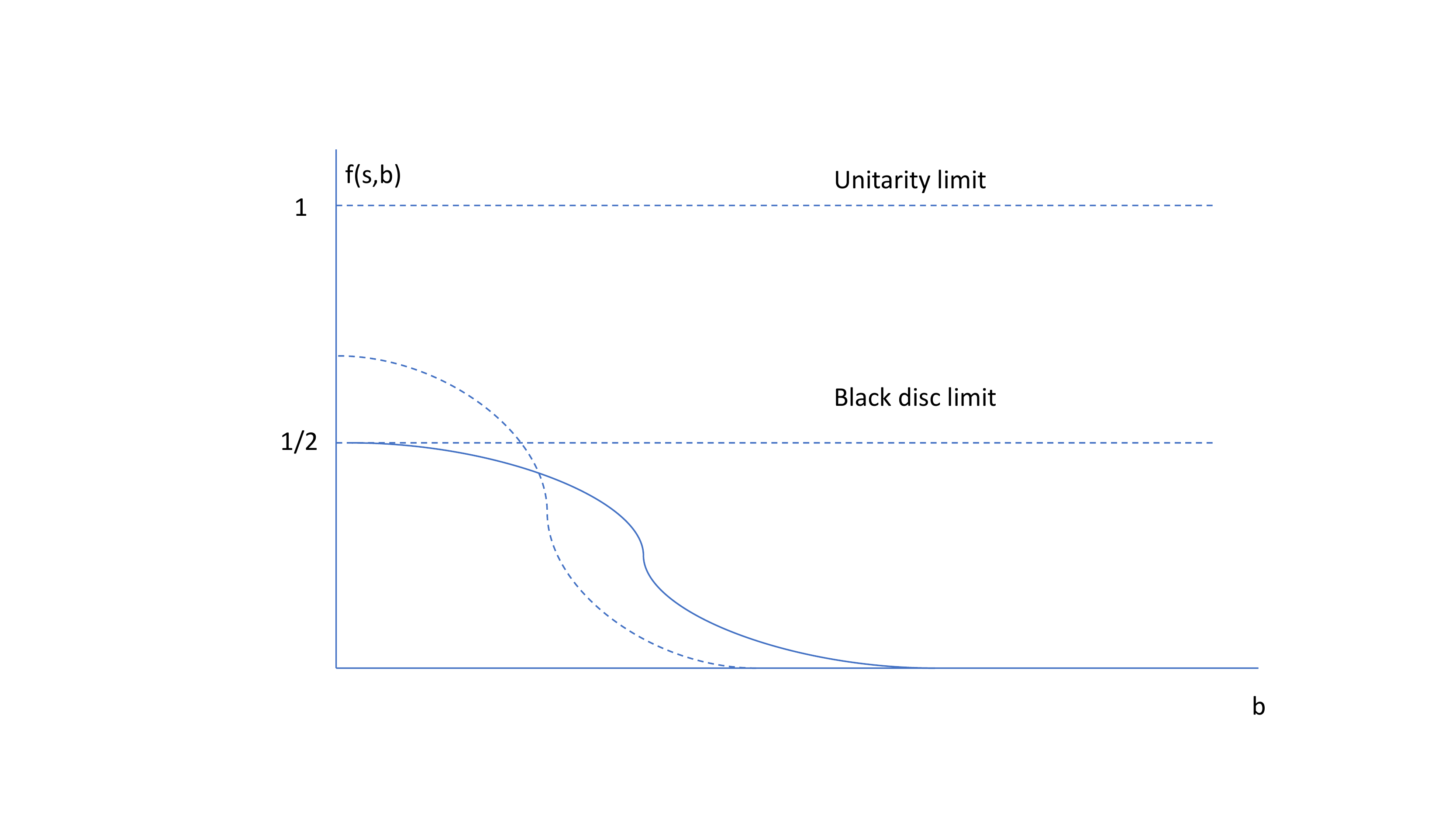}}
	\end{center}
	\vspace{-1cm}
	\caption[ch2]{\small Schematical impact-parameter dependence of the amplitude $f(s,b)$ in the two cases of fully absorptive  scattering (solid line) and partial transition to reflective  scattering (dashed line)  at the LHC highest energy \cite{new}.}
\end{figure}
The reason for coexistence of these different opinions  is that an excess over the black disc limit is  small even at the highest   LHC energy $\sqrt{s}=13$ TeV. However,
the numerical  impact parameter analysis performed in \cite{tsrg} has demonstrated that the effect of the black disc limit excess has  statistical significance more than $5\sigma$.

It should be emphasized that  the absorptive  models based on the standard eikonal unitarization  predict  an experimentally unobserved  appearance of the secondary bumps and dips  at large values of transferred momenta in the differential cross-section $d\sigma/dt$ \cite{mart}. It happens due to  flattening of the impact parameter form of the scattering amplitude at the LHC energies.

The distinctive feature of the transition to the reflective scattering mode is  developing of a peripheral form of the inelastic overlap function. This form starts to appear at the values of energy when the scattering amplitude $f$ becomes greater than $1/2$.
Here with $h_{inel}$ gets its maximum at some $b=r(s)$ instead of $b=0$, where $b=r(s)$ is  a solution of the equation $f(s,b)=1/2$ at the energy value $s$. 

Saturation of the black disc limit leads to the following limiting behavior of the functions: $X,Y\to 1/2$ and $Z\to 1/4$ at $s\to\infty$ . The limit for the function $Z(s)$ in the reflective scattering mode is $4$ times higher than the corresponding limit in the case of the black disk limit saturation. Such a difference of the asymptotic values for the function $Z(s)$ at $s\to\infty$ for the two scattering modes makes this function potentially important for  discrimination.  Theoretical argument by Chew and Frautchi on  maximality of asymptotic strength of strong interactions \cite{ch} is in favor of unity for the value of the effective interaction intensity $Y(s)$ as well as for the functions $X(s)$ and  $Z(s)$ at $s\to\infty$.  

Unfortunately, one can not expect that  experimental study of asymptotics   would become  possible in the near future. We are forced now  to study phenomena in the preasymptotic energy region. Does it mean that we can not obtain any information on the asymptotic dynamics? The answer to this question does not seem to be a  straightforward. The cosmic ray data have large error bars. Another option is to get a clue on the asymptotics from a direct reconstruction of the amplitude $f(s,b)$. Shortcomings of this approach have been mentioned already. 

Here we propose to use for that purpose   the experimental measurements at the available energies on   $Y(s)$ and its extrapolation for the higher energies considering the function $dY(s)/ds$ to detect starting point of its flattening. This energy  value correlates with  start of transition from a Gaussian--like form of the amplitude $f(s,b)$ to the Fermi--like dependence over the impact parameter. Fermi--like dependence being typical for nuclear reactions  produces secondary dips and bumps in the differential cross--section of the elastic proton scattering $d\sigma/dt$ at the asymptotic energies. Thus,  an apperance of the secondary dips and bumps  is an illuminating sign of  the asymptotics \cite{dips}.

The above signature is informative but due to  remotness of the asymptotics  is not  helpful at available energies. In this regard the old papers \cite{wall,wall1,strum}  being combined with  more recent ones  \cite{fm,fag,ben} are rather instrumental for  the numerical extrapolations of $Y(s)$.  Moreover, the interaction intensity  $Y(s)$ increases with energy faster than the ratio $X(s)$ and its further experimental measurements could provide  an earlier check of the asymptotics. This conclusion is also valid for the function $Z(s)$ which  is just a product of $X$ and $Y$. Comments on the energy dependence of  $Z(s)$ has been given above. Extrapolation of  $Z(s)$ to higher energies based on  \cite{fag}  is depicted on Fig. 3.
 \begin{figure}[hbt]
	\begin{center}
		\resizebox{12cm}{!}{\includegraphics*{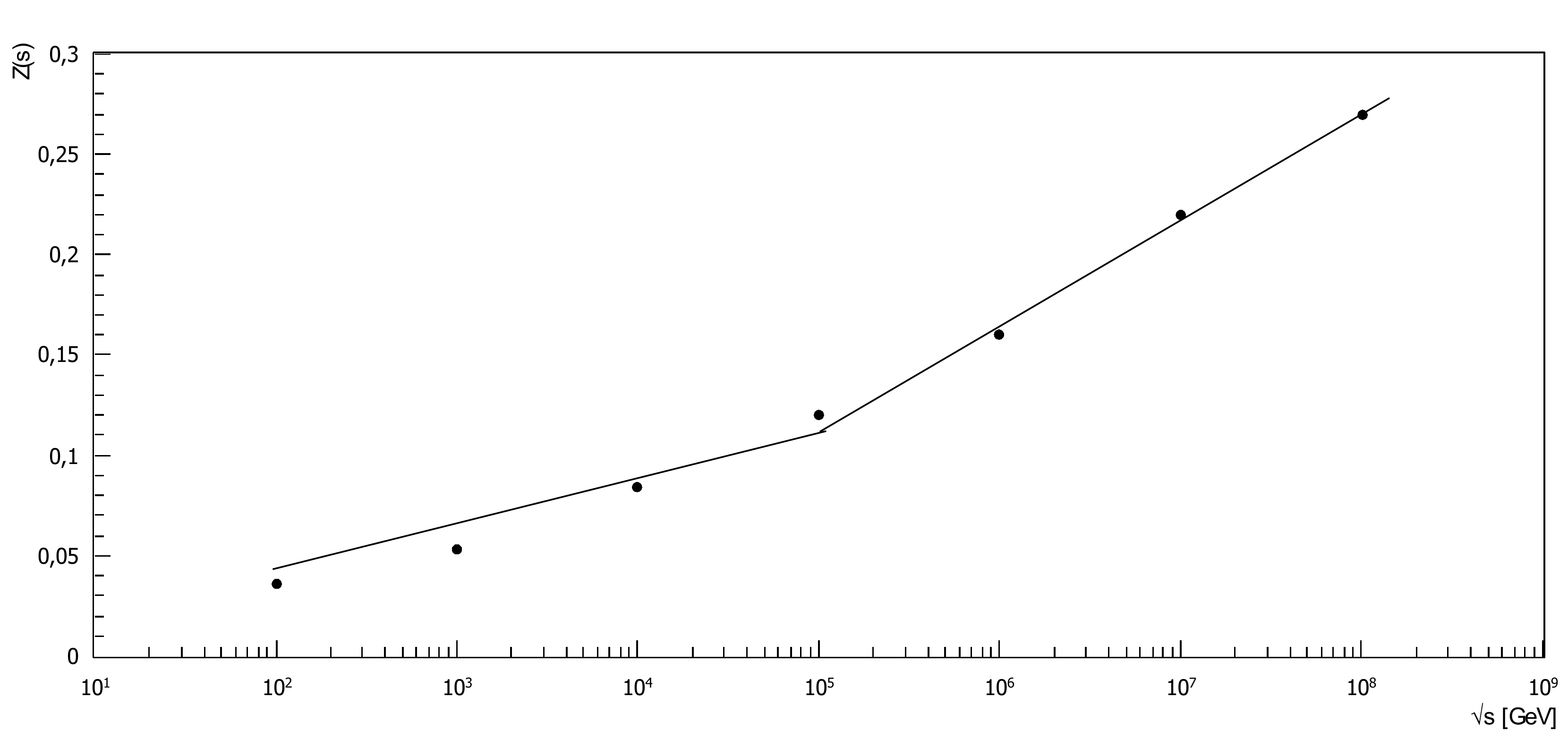}}
	\end{center}
	\caption[ch2]{\small Extrapolation of the energy dependence of the  function $Z(s)$ for the reflective scattering mode.}
\end{figure}
$Z(s)$ crosses the black disk limit $1/4$ at around $\sqrt{s}\simeq 10^8$ $GeV$. 
\section{Separating out the role of   interaction radius}
The energy dependencies of the experimental observables related to elastic scattering and soft hadron production are essentially affected by an interaction intensity behavior up to the LHC energies. Transition  to the asymptotic regime where the energy behavior of the observables are determined by the interaction radius only seems to lie in a rather remote energy range. The knowledge of the  value $Y(s)$ discussed here has practical implications and would be helpful under the study of  the energy range related to transition to the asymptotic regime and  relevant scattering mode. It would also allow  to separate out effects of increasing strength of interactions from the ones connected to an  increasing size of the interaction area. 

Such separation would be  helpful for understanding an origin of the total cross--section growth.  Simplified geometrical analogy when the effective total cross--section is proportional to the squared interaction radius, $\sigma_{tot}\propto R^2$,  should be corrected. This proportionality has nothing to do with the actual data obtained at the LHC but can only be accepted at the energies when the effective interaction intensity $Y(s)$ does not depend on $s$. It follows from a nontrivial energy dependence of  $Y(s)$ and representation of the total cross--section in the form
$$
\sigma_{tot}(s)=16\pi Y(s)B(s).
$$
Indeed, the effective interaction intensity measured experimentally increases by more than $50$$\%$ when $\sqrt{s}$ increases from $10^2$ to $10^4$ $GeV$. This estimate is based on the available experimental data for the ratio $X(s)$ and assumption on the approximate equality of the functions $X$ and $Y$.

The various dynamical mechanisms  proposed in the literature for explanation of  the total cross--section growth as an effect of   increasing interaction radius only are deficient ones at present energies where $Y(S)$ is not a constant and increase of the total cross--section is a combined effect of the increasing interaction intensity $Y(s)$ and of the interaction radius.   Significant difference among the observed energy dependencies of  $\sigma_{tot}(s)$  and the slope parameter $B(s)$ is in favor of the above conclusion. The major role of the interaction  radius  increase might be relevant in the asymptoical energy range, but not at the available energies. This fact should be taken into account under  model constructions pretending to describe the  effective cross-sections in the preasymptotic energy region.

\section{Conclusions}
We  suggested  definition of the effective interaction intensity for soft hadron interactions as a total interaction  strength averaged over size of the interaction area, i.e.
effective intensity  is determined by a strength of interaction per unit of the area of  the interaction region.
Such a definition  is useful for a description of   interactions at large distances since  soft hadron collisions provide a major contribution to the effective total cross--section. In this  region,  an explicit relation of the intensity of interaction with  running coupling constant of QCD has not been found due to unsolved problem of confinement and an existence of the spontaneous chiral symmetry breaking phenomenon.

The proposed discrimination of the scattering modes based on the  effective interaction intensity is a less sensitive method than the  straightforward reconstruction of the scattering amplitude $f(s,b)$, but on the other hand it does not require  of   additional assumptions. 

It should be noted that the ratio $X(s)=\sigma_{el}(s)/\sigma_{tot}(s)$  has been thoroughly discussed in \cite{fag} and, as it has been shown   $X(s)$ is approximately equal to  $Y(s)$.  This conclusion is definitely valid  for an exponential and step parameterizations of the $b$--dependence of $f(s,b)$, but not in the general case despite the
increase of the interaction intensity with energy is correlated with an increase of  $\sigma_{el}(s)/\sigma_{tot}(s)$ when $s$ grows. The essential increase of the function $X(s)$ has been observed  till the energy $\sqrt{s}=13$ $TeV$ and, as it has been noted above, could be interpreted as a result of a redistribution of the probabilities of elastic and inelastic collisions due to the reflective scattering mode appearance\cite{rin}. The elastic scattering increasingly decouples with energy from the inelastic events in this mode performing transition from shadow to geometric mechanism.

Detecting manifestations of asymptotics  is essential for studies of  hadron interaction dynamics. An expected asymptotic form of the scattering amplitude could correlate with general features  of a particular dynamics pointing out to its  stochastic or  coherent  nature. 

Related purpose of the effective interaction intensity introduction is a further clarification of the total cross--section growth origin.

\end{document}